# A Loop-Opening Model for the Intrinsic Fracture Energy of Polymer Networks

Shu Wang,[1,†] Chase M. Hartquist,[1,†] Bolei Deng,[1,†,‡] Xuanhe Zhao[1,2,]*

[1] Department of Mechanical Engineering, Massachusetts Institute of Technology, MA 02139, USA

[2] Department of Civil and Environment Engineering, Massachusetts Institute of Technology, MA 02139, USA

**ABSTRACT.** We present a loop-opening model that accounts for the molecular details of the intrinsic fracture energy for fracturing polymer networks. This model includes not only the energy released from the scission of bridging chains but also the subsequent energy released from the network continuum. Scission of a bridging chain releases the crosslinks and opens the corresponding topological loop. The released crosslinks will be caught by the opened loop to reach a new force-balanced state. The amount of energy released from the network continuum is limited by the stretchability of the opened loop. Based on this loop-opening process, we suggest that the intrinsics fracture energy per broken chain approximately scales with the product of the fracture force and the contour length of the opened loop. This model predicts an intrinsic fracture energy that aligns well with various experimental data on the fracture of polymer networks.

**Introduction.** Polymer networks are found in everyday items as diverse as automobile tires and hydrogels, but fracture events limit the lifespan of their service.[1–3] The connection between the fracture of polymer networks and their constituent polymer chains is unclear, hindering a deeper molecular understanding of their fracture behaviors. Lake and Thomas realized more than fifty years ago that the fracture energy of a polymer network is inherently linked to its polymer chains.[4] They defined the intrinsic fracture energy of a polymer network as the minimum energy needed to propagate a crack by creating a unit of new surface area, and assumed that the intrinsic fracture energy depends solely on the breaking of elastically active polymer chains. The Lake-Thomas model predicts the intrinsic fracture energy $\Gamma_0$ by multiplying the number of broken polymer chains per unit of new surface area $M$ with the energy needed to break a chain at the crack tip $U_{chain}$ (**Figure 1a**):

$$\Gamma_0 = MU_{chain} \tag{1}$$

The original Lake-Thomas model[4] further considers $U_{chain} = NU_{mon}$ as the energy for rupturing a bridging chain **(Figure 1a)**, where $N$ is the number of repeating units and $U_{mon}$ is the total bond dissociation energy per unit. Recently, Wang *et al.*[5] calculated the single-chain energy $U_{chain}$ as the area under the force-displacement curve up to the point where the chain fractures at the fracture force $f_f$, and showed that $U_{chain}$ is much lower than $NU_{mon}$. Wang *et al.* suggests that the reaction kinetics of chain scission dictate the criterion for crack propagation.[5] With their estimation, the intrinsic fracture energy predicted by Equation 1 underestimates various experimental results by 1 to 2 orders of magnitude (**Table 1**).

**Table 1.** Comparison between $\Gamma_0/M$ and corresponding $U_{chain}$ of end-linked poly(ethylene glycol) (PEG) networks.[6–9]

|  | Wang et al. | Lin et al. | Akagi et al. | Barney et al. |
| --- | --- | --- | --- | --- |
| $\Gamma_0/M$ [a)] (nN·nm) | 430 | 872 | 351 | 528 |
| $U_{chain}$ [b)] (nN·nm) | 5.5 | 14.7 | 4.6 | 3.6 |
| $\Gamma_0/MU_{chain}$ | 78 | 60 | 76 | 145 |

[a)] Areal number density $M$ is estimated based on the storage moduli of the networks; [b)] $U_{chain}$ is estimated based on the data of storage moduli, single-molecule force spectroscopy experiments, and typical breaking forces of the backbone bonds. Detailed calculations can be found in the supporting information.

Extensive efforts have been made to explain why the Lake-Thomas model significantly underestimates the intrinsic fracture energy of polymer networks.[6,9–11] Several tree-like models[10,11] have been proposed to account for the energy released from unbroken chains connecting to a bridging chain when it breaks at $f_f$ (**Figure 1b**). These models suggest that $\Gamma_0/M$ should be equated with the energy stored in the tree-like structure rather than the single-chain energy. The tree-like models predict a higher intrinsic fracture energy per bridging chain, especially when considering infinite trees.[11] However, the concept of infinitely large trees does not realistically apply to actual polymer networks, since real polymer networks consist of finite trees due to finite topological loops.[12,13]

Numerical simulations of lattice-like polymer network models have demonstrated that $\Gamma_0/M$ is inherently much larger than $U_{chain}$.[14,15] The lattice-like polymer networks have well-defined topologies (**Figure 1c**), where each junction is freely jointed with the same functionality. Each network edge exhibits the typical force-extension behavior of synthetic polymer chains, with scission governed by the fracture force $f_f$ along the edge. The intrinsic fracture energies in these networks were measured using the "Rivlin-Thomas" method in the pure-shear geometry.[1] Comparisons between the simulated $\Gamma_0$ and the predictions of the Lake-Thomas model (Equation 1) revealed that the intrinsic fracture energy obtained by fracturing the network is significantly larger than $MU_{chain}$, consistent with the data in **Table 1**. The numerical simulations of the lattice-like polymer network models suggest that a substantial amount of energy is released and dissipated within the network continuum. While the lattice-like network unveils that the nonlocal energy dissipation in the network continuum significantly contributes to the intrinsic fracture energy, it is obviously an oversimplification (e.g., **Figure 1c**) for the topology of realistic polymer networks.

Here we propose a loop-opening model for the intrinsic fracture energy of polymer networks that accounts for both nonlocal energy dissipation in the network continuum and the polymer network topology (**Figure 1d**). This model includes finite topological loops that open during crack propagation, unlike the infinite topological loops in the tree-like models. The retraction of the network continuum (shaded in **Figure 1d**) balances with the unravelling of the loop (blue strands in **Figure 1d**), and the new force-balanced state is largely dictated by the stretchability of the opened loop. The loop-opening model provides a new scaling law for the intrinsic fracture energy of polymer networks,

$$\frac{\Gamma_0}{M} \sim f_f L_f \qquad (2)$$

where $f_f$ and $L_f$ are the critical force and displacement at the chain fracture. The loop-opening model (Equation 2) predicts the intrinsic fracture energy to be one to two orders of magnitude higher than the Lake-Thomas model (Equation 1).

In the remainder of this paper, we will introduce the force-extension behavior of a single polymer chain and the corresponding network continuum and discuss the concepts and limitations of existing fracture models of polymer networks. We will then integrate the single-chain behavior with the network continuum to demonstrate our proposed loop-opening model and its new scaling law. At the end, we will validate our model by comparing it with experimental data.

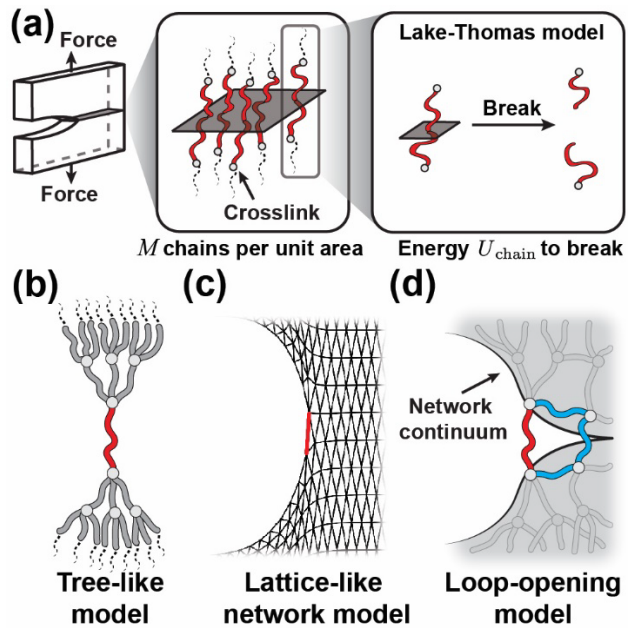

**Figure 1.** (a) Schematic illustration of various models for the intrinsic fracture energy of polymer networks. The Lake-Thomas model only considers the energy released and dissipated in the bridging chains. (b) The tree-like model considers the energy released and dissipated in the network continuum as Caley tree structures. (c) The lattice-like network model is a numerical model that considers the energy released and dissipated in the network continuum, but it simplifies the networks to well-defined lattice structures. (d) The loop-opening model of polymer network considers the opening of finite topological loops and the retraction of the network continuum after bridging chain scission. It considers both the energy released and dissipated in the network continuum.

**Single chain and network continuum.** We begin with the force-extension behavior of single polymer chains.[16,17] Experiments have shown that the force-extension behaviors of common synthetic polymers align well with the modified freely-jointed chain model (m-FJC), [16,17]

$$\frac{R(f)}{L} = \mathcal{L}\left(\frac{f}{f_s}\right)\left(1 + \frac{f}{f_e}\right) \qquad (3)$$

where $R$ is the end-to-end distance, $f$ is the tension along the chain, $L$ is the force-free contour length of the polymer, and $\mathcal{L}(x) = \coth(x) - 1/x$ is the Langevin function. The characteristic entropic tension $f_s = kT/b$ characterizes the initial linear entropic elasticity at low force regime, where $k$ is the Boltzmann constant, $T$ is the absolute temperature, and $b$ is the Kuhn length. The characteristic enthalpic tension $f_e$ characterizes the linear enthalpic elasticity, which is typically three orders of magnitude larger than $f_s$. It describes the linear extension of the polymer backbone beyond force-free contour length $L$. Note that the polymer chain always breaks at $f_f < f_e$. The energy stored in a bridging chain at breakage $U_{chain}$ is the area under the force-displacement curve up to the fracture point (**Figure 2a**).

When the crack propagates in a polymer network, a bridging chain is not directly connected to rigid boundaries. Instead, it is connected to an elastic network continuum that consists of other polymer chains (**Figure 1d**). We could simplify this by imagining a polymer chain is connected to rigid boundaries through two network continua (**Figure 2b**). To reach the fracture force of the bridging polymer chain $f_f$, the boundaries need to be displaced to $L_{boundary}$, which is much longer than the fracture length of the bridging chain $L_f$. At the time the bridging chain ruptures, the network continua store a significant amount of energy (grey area in **Figure 2b**). Additionally, the force-displacement curve for this whole combination (**Figure 2b**) is much more compliant than the single bridging chain at high forces. Because the load in the network continua is shared by many polymer chains in the direction of deformation, allowing the network continua to act like more compliant springs that stores significant amount of energy.

As an example, we can approximate the network continua as two Caley trees connected to each end of the bridging chain, comprising identical polymer chains and crosslinks (structures shown in **Figure 2c**). The trees' roots are linked to rigid boundaries. As the rigid boundaries separate, the tree-like structure deforms and stores elastic energy. With identical chains, the tensions and end-to-end displacements of all polymer chains in the same generation of the tree-like structure (**Figure 2c**) are almost the same. The tension of the bridging chain (shown in red in **Figure 2c**) equals the sum of the tensions of all chains in one generation of the tree-like structure. Therefore, the tension of a polymer chain in the $g$-th generation of the tree-like structure is $f/(z-1)^{|g|}$, where $z$ is the functionality of crosslinks.[18] Applying Equation 3 to a tree-like structure, the force-displacement relationship of a tree-like structure can be expressed as[11]

$$R_{tree}(f) = \sum_g R\left(\frac{f}{(z-1)^{|g|}}\right) = L \cdot \sum_g \mathcal{L}\left(\frac{f/(z-1)^{|g|}}{f_s}\right)\left(1 + \frac{f/(z-1)^{|g|}}{f_e}\right) \quad (4)$$

where $R_{tree}$ is the displacement of the boundary that the trees' roots are connected with, $f$ is the tension on the bridging chain in $g = 0$ (or the combined tension in a generation). A typical force-displacement curve obtained by Equation 4 is shown in **Figure 2c**. Compared to a single chain, the tree-like structure deforms compliantly at high forces, extends much further, and stores more energy before the bridging chain ruptures. Although the scheme demonstrated here does not fully mirror the scenario of the bridging chain at the crack tip, this approximation qualitatively captures two key features: 1) the bridging chain at $g = 0$ has the highest tension (analogous to stress concentration at the crack tip of loaded notched polymer networks); 2) the network continuum stores a lot of energy at the moment the bridging chain ruptures.

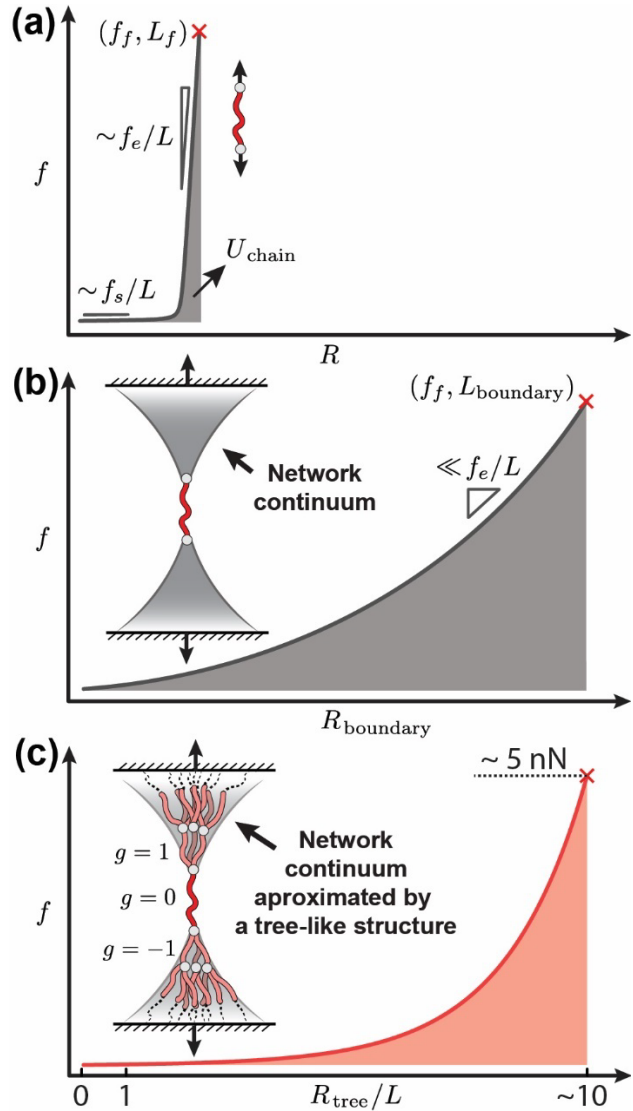

**Figure 2.** Schematic force-displacement curves for (a) a polymer chain, and (b) a polymer chain connected to the rigid boundaries through elastic network continua. (c) a polymer chain connected to the rigid boundaries through tree-like structures as the elastic network continua. Typical polymer chains with $f_f \approx 5$ nN, $f_s \approx 6$ pN and $f_e \approx 100$ nN (data from PEG chains[17,19]) provide the tree-like structure with $z = 4$ a rupture displacement about 10 times that of the bridging chain contour length $L$.

**The loop-opening model.** Real polymer networks consist of finite topological loops.[12,13,20] The scission of a bridging chain leads to the opening of the corresponding loop. Consider a defect-free notched network comprised of identical polymer chains (e.g., no dangling chains and 1st or 2nd order loops). At the crack tip, we may consider that the network continua (shaded in **Figure 3a**) connect the bridging chain (dark gray in **Figure 3a**) to the distant boundaries, with a topological loop (blue) situated ahead of the bridging chain (red). When the bridging chain is quasi-statically loaded to the fracture force $f_f$, the displacement of the two crosslinks on the bridging chain $R$ approaches the fracture displacement $L_f$ (transitioning from state 1 to state 2 in **Figure 3a**). Throughout this process, the restoring force along the bridging chain can be represented by the m-FJC model (gray curve in **Figure 3c**). Meanwhile, the network continua store elastic energy as well. When the bridging chain fractures, its two crosslinks are left highly unbalanced due to the force $f_f$ exerted by the elastic network continua (state 2 in **Figure 3a**). These crosslinks then seek a new force-balanced state (state 3, **Figure 3a**). While the process from state 2 to state 3 is complex, we can still estimate the energy released.

Approximating the elastic network continua as the tree-like structures without the loss of generality (red chains in **Figure 2c&3a**), we then decompose the process from state 2 to state 3 in **Figure 3a** into two components: the retraction of the elastic network continua (red tree-like structures in **Figure 3b**) and the extension of the opened loop (blue in **Figure 3b**). We track these contributions by monitoring the imaginary forces required to quasi-statically unload the elastic network continuum is $f_{tree}$ (Equation S5), while the imaginary force that is required to quasi-statically load the opened loop is $f_{loop}$. These two forces are plotted in **Figure 3c** in red and blue, respectively. The former is much larger than the latter at state 2 ($f_{tree} \gg f_{loop}$), but they counterbalance at state 3 ($f_{tree} = f_{loop}$). In this assumed quasi-static unloading process, the energy released from state 2 to state 3 is $U_{release}$ (red shaded area in **Figure 3c**), and the crosslinks are displaced from $L_f$ to approximately ($n_{loop} - 1$) $L_f$ (**Figure 3c**), where $n_{loop}$ is the number of polymer chains within the loop.

The energy released from state 2 to state 3 is eventually dissipated in the polymer network (e.g., viscous dissipation), which contributes to the measured intrinsic fracture energy of the polymer network. Therefore, the loop-opening model predicts $\Gamma_0/M \approx U_{chain} + U_{release} \approx U_{release}$, as $U_{release} \gg U_{chain}$. This energy $U_{release}$ depends on the retraction behavior of the elastic network continua (red curve) and the stretchability of the opened loop (blue curve).

Although we use the tree-like structures to approximate the network continua, it should be noted that the loop-opening model does not depend on the tree-like structures for the elastic network continua. Instead, any constitutive models that can characterize the generic features of the elastic network continua (**Figure 2b**) are applicable to the loop-opening model. Additionally, the opened loop should balance the retraction of the network continuum, and its stretchability determines a limit for the retraction (**Figure 3b&c**). Therefore, rather than deriving an analytical solution for $U_{release}$, we propose a semi-quantitative argument that $U_{release}$ scales with $f_f \cdot (n_{loop} - 2) L_f$ (**Figure 3d**). This represents the area of the rectangle between states 2 and 3 in **Figure 3d**, with ($n_{loop} - 2$) $L_f$ as the displacement of two crosslinks from state 2 to state 3 and $f_f$ as the typical force scale during this process. Based on the loop-opening model, we can thus formulate a new model for the intrinsic fracture energy of polymer networks,

$$\frac{\Gamma_0}{M} = \alpha f_f L_f \tag{5}$$

where parameter $\alpha$ is a parameter that scales with $n_{loop} - 2$.

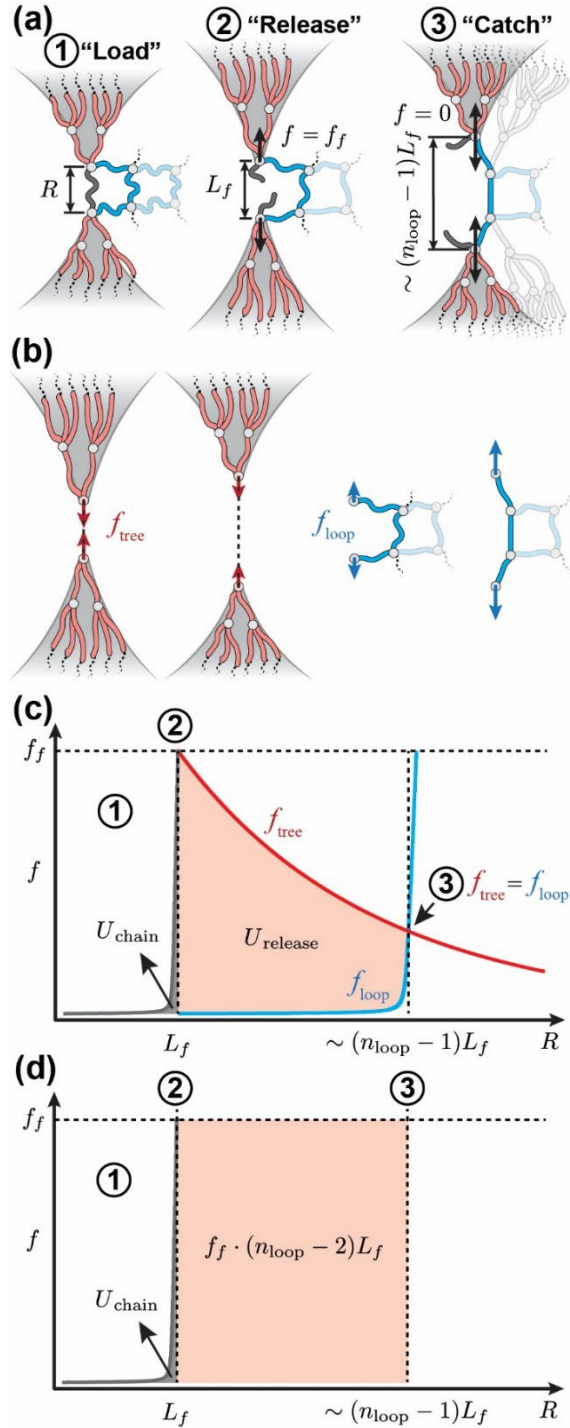

**Figure 3.** (a) Schematic illustration of the loop-opening model that consists of a three-step process: load, release, then catch. State 2 is the moment when the bridging chain just breaks, leaving the network continua unstable. State 3 is the moment when the crosslinks of the broken bridging chain reach a new force-balanced condition. (b) From State 2 to State 3, we decompose this process into two components: the retraction of the network continua (red trees) and the extension of the opened loop (blue chains). (c) The imaginary force that quasi-statically unloads the tree is shown in red, and the imaginary force that quasi-statically loads the opened loop is shown in blue. he difference gives the energy released in this process as $U_{release}$. (d) Direct comparison between the single-chain energy $U_{chain}$ considered by the Lake-Thomas model and the energy $f_f \cdot (n_{loop} - 2) L_f$ considered in the semi-quantitative loop-opening model.

**Comparison with experimental results.** We next validate the loop-opening model by comparing its prediction of the intrinsic fracture energy with experimental results. The experiments were carried out by various groups using end-linked poly(ethylene glycol) (PEG) networks. The networks were synthesized through different end-linking strategies, including the end-linking of two different tetra-arm PEG macromers ($A_4+B_4$) by Akagi *et al.* and Lin *et al.*, the end-linking of a tetra-arm PEG macromer with bifunctional small-molecule linkers ($A_4+B_2$) by Wang *et al.*, the end-linking of linear PEG with tetra-functional small-molecule crosslinkers ($A_2+B_4$) by Barney *et al.* and Arora *et al.*[7–9,21,22] The gels were prepared under semi-dilute conditions and well below the entanglement concentration to avoid trapped entanglements. The moduli therefore correspond directly to the contributions of elastically active chains. The moduli of the gels have been confirmed to be relatively consistent with the phantom network model and the real elastic network theory model. Therefore, the information about the elastically active chains, such as the areal number density $M$ and the single-chain energy $U_{chain}$, can be extracted from the moduli of these gels. Detailed calculation can be found in the supporting information. The experimental results of $\Gamma_0/M$ from various groups are plotted against $f_f L_f$ in **Figure 4** with red circles. The loop-opening model with different pre-factors $\alpha$ (solid red lines in **Figure 4**) are plotted alongside. The model and experimental data align well, and both significantly exceed the predictions of the Lake-Thomas model (Equation 5 and black dots in **Figure 4**) and the tree-like model (blue dots in **Figure 4**).

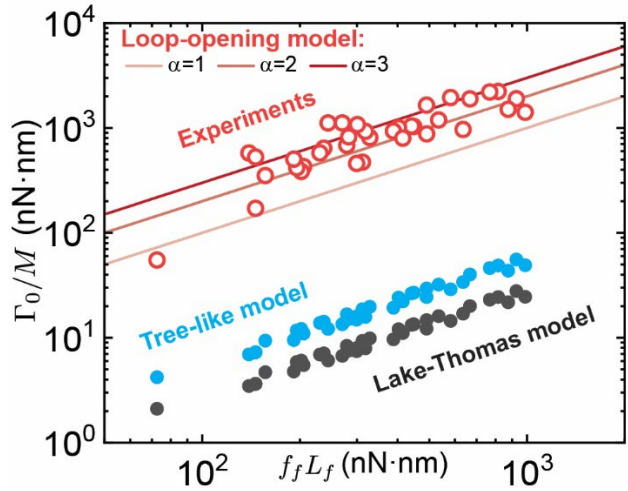

**Figure 4.** Plot of different fracture energy per chain $\Gamma_0/M$ against corresponding $f_f L_f$. Solid lines depict the scaling law $\Gamma_0/M = \alpha\, f_f\, L_f$ with $\alpha = 1, 2, 3$. Red circles are $\Gamma_0/M$ estimated from experimental data of end-linked PEG gels. Black dots are $\Gamma_0/M = U_{chain}$ based on the Lake-Thomas model. Blue dots are $\Gamma_0/M \approx [z/(z-2)]\, U_{chain} = 2U_{chain}$ based on the tree-like model[10,11] with functionality $z = 4$, assuming small loop[12,13].

**Discussion.** Several points regarding the loop-opening model are noteworthy. First, various models for the intrinsic fracture energy of polymer networks (**Figure 1**) assume different length scales for energy dissipation. The Lake-Thomas model only considers the energy dissipated in the bridging chain. The tree-like model considers the energy that can be dissipated in a tree-like structure. This involves the contribution from unbroken chains, but it still considers the energy to be dissipated below the scale of topological loops, as the tree-like structure only exists within the scale of such loops.[12,13] Numerical simulations of the lattice-like network model reveal the energy can be released and dissipated much beyond the scale of topological loops.[14,15] Albeit oversimplification, it leads to the physical picture developed in the loop-opening model which considers the interplay between network continuum and the broken bridging chain.

Second, for the same type of polymer networks, the loop-opening model (Equations 2&5) leads to

$$\frac{\Gamma_0}{M} \sim N \qquad (6)$$

since the fracture force $f_f$ remains constant while $L_f \sim N$, where $N$ is the number of Kuhn segments of the network chains between crosslinks. This mirrors the scaling law of the Lake-Thomas model (Equation 1), since $U_{chain} \sim N$. The difference is that the estimated values of $\Gamma_0$ from the loop-opening model are approximately two orders of magnitude larger than that from the Lake-Thomas model.

Third, for polymer networks prepared under highly overlapped conditions,[12,23] overlapped topological loops and entanglements might impede complete loop-opening, leading to potential deviations from the loop-opening model (Equation 5).

Fourth, the loop-opening model for polymer networks shares some similarity with the lattice trapping model in crystal fracture. Both models focus on the local stress and load at the crack tip, extending beyond the thermodynamic limit of the Griffith theory[24] based on bond dissociation energy. The former is pronounced in polymer networks according to **Table 1** and **Figure 4**, while the latter is almost experimentally inaccessible in crystal due to thermal fluctuation. For more details on the lattice trapping model, readers are referred to the works of Thomson *et al.*[25], Curtin[26], and Marder[27].

**Conclusion.** The loop-opening model introduced in this study captures the intrinsic fracture energy of polymer networks by focusing on the energy released from network continuum. The energy released per broken chain, $\Gamma_0/M$, is governed by the stretchability of polymer chains within the opened loop. Without further assumption for the constitutive law of the network continuum, this leads to a new relationship, $\Gamma_0/M = \alpha\, f_f L_f$, for polymer network fracture, which aligns well with experimental data from end-linked PEG gels. The model suggests that the intrinsic fracture energy of a polymer network stems from contributions at both the single-chain level and the network-continuum level. Moreover, the loop-opening model suggests that by either shielding the bridging chain from the fracture force or increasing its crosslinks' maximum displacement before breaking the next chain (e.g., through larger topological loops) could enhance the fracture energy, resulting in tougher and more resilient polymeric materials. Coupled with continuous advancements in polymer characterization technology and precise chemical control of the force-coupled reactivities,[7,28,29] the topologies of polymer networks,[30,31] and chain conformations,[32] our model could offer deeper insights into the interplay between network fracture and the mechanics of individual chains, thereby paving the way for the design of more robust polymeric materials.

**Supporting Information**.

Analysis of the experimental data in Table 1 and Figure 4; Dependence of force-displacement on the retraction of tree-like structures.

AUTHOR INFORMATION

**Corresponding Author**

*Send correspondence to zhaox@mit.edu

**Present Addresses**

‡ Daniel Guggenheim School of Aerospace Engineering Georgia Institute of Technology, Atlanta, Georgia 30332, USA

**Author Contributions**

† Co-first authors.

The manuscript was written through contributions of all authors. All authors have given approval to the final version of the manuscript.

**Funding Sources**

This work is supported in part by the National Institutes of Health (Grants No. 1R01HL153857 01 and No. 1R01HL167947-01), the National Science Foundation (Grant No. EFMA-1935291), and Department of Defense Congressionally Directed Medical Research Programs (Grant No. PR200524P1).

**Notes**

The authors declare no competing financial interest.

ACKNOWLEDGMENT

The authors thank W.A. Curtin for helpful discussion.

For Table of Contents use only

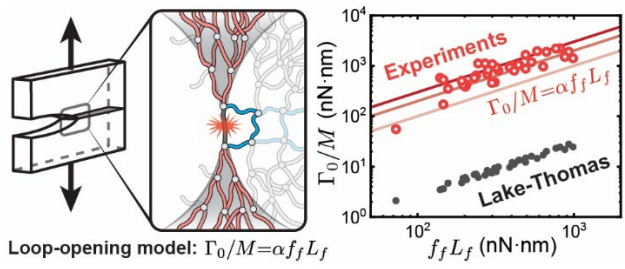

# A Loop-Opening Model for the Intrinsic Fracture Energy of Polymer Networks

Shu Wang, [1,†]  Chase M. Hartquist, [1,†]  Bolei Deng, [1,†,‡]  Xuanhe Zhao[1,2,]*

[1]Department of Mechanical Engineering, Massachusetts Institute of Technology, MA 02139, USA

[2]Department of Civil and Environment Engineering, Massachusetts Institute of Technology, MA 02139, USA

**Experimental data in Table 1 and Figure 4.** The data in Table 1 and Figure 4 come from experiments by Wang *et al.*, Lin *et al.*, Akagi *et al.*, Barney *et al.*, and Arora *et al.* on end-linked poly (ethylene glycol) (tetra-PEG) gels.[1–5] Wang *et al.* prepared gels using 5 kDa tetra-PEG and a copper-catalyzed alkyne azide coupling (CuAAC) click reaction, with $A_4+B_2$ end-linking (where $A_4$ is tetra-arm PEG with azide groups and $B_2$ is bi-functional alkyne linkers).[1] These linkers designed based on mechanophores control the fracture force $f_f$ of the polymer backbone. The gels, formed just above their overlap concentration, had similar structures but varied in strand fracture force. Akagi *et al.* and Lin *et al.* used different arm-length tetra-PEGs (5, 10, 20, 40 kDa) and prepared the gels under different preparation volume fractions ($\phi_0$ = 0.034, 0.05, 0.066, 0.081, 0.096), employing $A_4+B_4$ end-linking (with $A_4$ and $B_4$ representing tetra-arm PEG with amine and N-Hydroxysuccinimide ester functional groups, respectively).[2,3] Barney *et al.* used linear-PEGs with different molecular weight (4, 8, 12 kDa) and prepared the gels under different preparation concentrations ($c$ = 50, 62, 77, 100 g/L) by end-linking the linear-PEGs with tetra-arm small molecule crosslinkers ($A_2$ +

B$_4$).[4] Arora *et al.* used 5 kDa linear-PEGs and prepared the gels under different preparation concentrations ($c$ = 8, 12, 16, 18 mM) by end-linking the linear-PEGs with tetra-arm small molecule crosslinkers (A$_2$ + B$_4$).[5] Fracture energies from these experiments, measured via tearing tests under steady-state uniaxial extension and cyclic loading, are all considered as the intrinsic fracture energies of the gels. This is because tetra-PEG gels have trivial viscous dissipation, which makes the data from steady-state extension fracture tests and cyclic fatigue tests very similar.[2]

To estimate the intrinsic fracture energy per strand $\Gamma_0/M$, we estimate the areal number density $M$ of elastically active strands of each gel by $M \approx \frac{1}{2} \nu R_0$,[6] where $R_0 \approx bN^{0.588}$ is the root-mean-square end-to-end distance of the elastically active strands in the undeformed state ($b$ is the Kuhn length and $N$ is the number of Kuhn segments). The real chain scaling is applied based on the assumption that strands are at or slightly above overlap under the preparation condition and that the solvent is a good solvent, like water[7] (note that Wang *et al.*[1] and Arora *et al.*[5] prepared their gels in propylene carbonate, and Barney *et al.*[4] prepared their gels in dimethyl formaldehyde). The average number of the Kuhn segments between junctions $N$ can be estimated based on the phantom model[8] of network elasticity (functionality = 4) from the shear modulus $G$, assuming the reaction extent of cross-linking is 100%,

$$G = (\nu - \mu)kT = \frac{1}{2}\nu kT = \frac{1}{2}\frac{cRT}{NM_0} \tag{S1}$$

where $\nu$ is the number density of elastically active strands, $\mu$ is the number density of elastically active junctions ($\mu = \nu/2$ when functionality is 4), $k$ is the Boltzmann constant, $T$ is absolute temperature, $c$ is the mass concentration of the polymer, and $M_0$ is the molecular weight of a Kuhn segment. Given the $M_0$ of PEG is 137 g/mol and the Kuhn length $b$ is 1.1 nm,[9] $N$, $R_0$ and $M$ can thus be estimated based on the shear modulus.

Estimating the single-chain energy $U_{chain}$ and $f_f L_f$ requires the fracture force $f_f$ and the force-extension relationship for individual polymer strands. Note that the forces $f_f$ of gels by Wang *et al.* were controlled by the mechanophores embedded and the potential weak bonds along the polymer backbone, which were estimated to be 1.3, 2.6, 3.8 nN, respectively.[1] The forces $f_f$ of gels by Lin et al.[2] and Akagi et al.[3] are considered to be close to 5 nN since all chemical bonds on the backbone are similarly strong as common C-C and C-O bonds.[10] The forces $f_f$ of gels made by Barney et al.[4] are considered to be around 2.5 nN, since they contain C-S backbone bond as preferential scission site.[11] The forces $f_f$ of gels made by Arora et al.[5] are considered to be around 3.8 nN, since the alpha C-C bond of the triazole group is a preferential scission site.[12]

We use the modified freely jointed chain model (m-FJC) (Equation 3) to describe the relationship between the force $f$ and the end-to-end distance $R$ of the polymer chain. The entropic $f_s$ and energetic $f_e$ characteristic tensions determine the compliant entropic elasticity at the low force regime and the stiff energetic elasticity at the high force regime, respectively. For PEG polymers, $f_s = 5.86$ pN and $f_e = 105$ nN.[13] The single-chain energy $U_{chain}$ is calculated for each gel using the equation below[14]:

$$U_{chain} = Lf_s \left\{ \left(\frac{f_f}{f_s}\right)\left[\coth\left(\frac{f_f}{f_s}\right) - \frac{f_s}{f_f}\right] + \ln\left[\frac{\frac{f_f}{f_s}}{\sinh\left(\frac{f_f}{f_s}\right)}\right] + \frac{f_f^2}{2f_s f_e} \right\} \quad (S2)$$

For $f_f \gg f_s$ (synthetic covalent polymers usually have $f_f / f_s \approx 1000$), an asymptotic form can be used,

$$U_{chain} \simeq Lf_s \left\{ \left(\frac{f_f}{f_s}\right)\left[\coth\left(\frac{f_f}{f_s}\right) - \frac{f_s}{f_f}\right] + \left[\ln\left(\frac{f_f}{f_s}\right) - \frac{f_f}{f_s} - \ln\frac{1}{2}\right] + \frac{f_f^2}{2f_s f_e} \right\} \quad (S3)$$

The product $f_f L_f$ can be calculated using the following equation based on Equation 3:

$$f_f L_f = f_f L \left[\coth\left(\frac{f_f}{f_s}\right) - \frac{f_s}{f_f}\right]\left(1 + \frac{f_f}{f_e}\right) \quad (S4)$$

The average contour length of elastically active strands within each PEG network are estimated ($L = Nb = N \times 1.1$ nm), where $N$ can be calculated from the shear moduli (Equation S1). Combining $L$ for different gels with the fracture force $f_f$ and parameters $f_s$ and $f_e$ from before, $U_{chain}$ and $f_f L_f$ can be estimated using Equation S3 and S4.

Table S1. Summary of data of A$_4$ + B$_4$ tetra PEG networks by Akagi et al.[3] and Lin et al.[2] used in Figure 4.

| MW (kDa) | $c$ [a] (g/L) | $G$ [b] (kPa) | $\Gamma_0$ [b] (J/m²) | $N$ [c] | $L$ [d] (nm) | $U_{chain}$ [e] (nN·nm) | $f_f L_f$ [e] (nN·nm) | $\Gamma_0/M$ (nN·nm) |
|---|---|---|---|---|---|---|---|---|
| 5 | 56.3 | 9.4 | 12.4 | 54 | 60 | 9.3 | 312 | 472 |
| 5 | 74.3 | 19.1 | 16.0 | 35 | 39 | 6.1 | 202 | 385 |
| 5 | 91.1 | 24.3 | 21.5 | 34 | 37 | 5.9 | 196 | 417 |
| 5 | 108.0 | 35.9 | 23.5 | 27 | 30 | 4.6 | 157 | 351 |
| 10 | 56.3 | 8.9 | 20.8 | 57 | 63 | 9.8 | 329 | 808 |
| 10 | 74.3 | 13.8 | 24.8 | 48 | 53 | 8.4 | 279 | 683 |
| 10 | 91.1 | 20.0 | 30.5 | 41 | 45 | 7.1 | 237 | 640 |
| 10 | 108.0 | 24.4 | 33.1 | 40 | 44 | 6.9 | 231 | 578 |
| 20 | 56.3 | 5.5 | 25.0 | 93 | 102 | 16.0 | 535 | 1189 |
| 20 | 74.3 | 8.6 | 30.8 | 78 | 86 | 13.4 | 448 | 1031 |
| 20 | 91.1 | 10.7 | 38.4 | 77 | 84 | 13.2 | 442 | 1041 |
| 20 | 108.0 | 14.0 | 45.6 | 70 | 77 | 12.1 | 402 | 1004 |
| 40 | 56.3 | 3.2 | 31.8 | 161 | 177 | 27.8 | 928 | 1894 |
| 40 | 74.3 | 4.7 | 51.9 | 142 | 156 | 24.4 | 816 | 2220 |
| 40 | 91.1 | 6.2 | 64.7 | 134 | 147 | 23.0 | 769 | 2201 |
| 40 | 108.0 | 8.4 | 69.7 | 116 | 128 | 20.0 | 668 | 1889 |
| 20 | 100 | 10.6 | 34.0 | 85 | 94 | 14.7 | 491 | 872 |

[a] Concentrations are the product of volume fraction $\phi_0$ and density of PEG (1.125 g/mL). [b] The data have been estimated from the plots by Akagi et al.[3], and certain errors exist due to the estimation process. [c] Average number of Kuhn segments in an elastically active chain is estimated based on Equation S1. [d] Force-free contour length $L = bN$. [e] Fracture force for a polymer chain is considered to be 5 nN.

**Table S2.** Summary of data of $A_4 + B_2$ tetra PEG network by Wang et al.[1] used in Figure 4.

| MW (kDa) | c (g/L) | G (kPa) | $\Gamma_0$ (J/m²) | N [a] | L [b] (nm) | $f_f$ (nN) | $U_{chain}$ (nN·nm) | $f_f L_f$ (nN·nm) | $\Gamma_0/M$ (nN·nm) |
|---|---|---|---|---|---|---|---|---|---|
| 5 | 128 | 23.0 | 3.4 | 50 | 55 | 1.3 | 2.1 | 72 | 55 |
| 5 | 128 | 23.2 | 10.6 | 50 | 55 | 2.6 | 3.6 | 142 | 171 |
| 5 | 128 | 24.4 | 27.1 | 47 | 52 | 3.8 | 5.5 | 199 | 430 |

[a] Average number of Kuhn segments in an elastically active chain is estimated based on Equation S1. [b] Force-free contour length $L = bN$.

**Table S3.** Summary of data of $A_2+B_4$ linear PEG network by Barney et al.[4] and Arora et al.[5] used in Figure 4.

| MW (kDa) | c (g/L) | G [a] (kPa) | $\Gamma_0$ [a] (J/m²) | N [b] | L [c] (nm) | $U_{chain}$ [d] (nN·nm) | $f_f L_f$ [d] (nN·nm) | $\Gamma_0/M$ (nN·nm) |
|---|---|---|---|---|---|---|---|---|
| 4 | 50 | 4.3 | 8.1 | 105 | 115 | 7.4 | 299 | 459 |
| 4 | 62 | 8.3 | 13.3 | 68 | 74 | 4.8 | 191 | 503 |
| 4 | 77 | 14.0 | 21.5 | 50 | 55 | 3.5 | 139 | 576 |
| 4 | 100 | 17.4 | 25.1 | 52 | 57 | 3.6 | 146 | 528 |
| 8 | 50 | 3.3 | 15.0 | 136 | 149 | 9.6 | 388 | 934 |
| 8 | 62 | 5.0 | 19.8 | 113 | 125 | 7.9 | 318 | 927 |
| 8 | 77 | 7.2 | 31.7 | 96 | 106 | 6.7 | 270 | 1115 |
| 8 | 100 | 10.4 | 42.9 | 87 | 96 | 6.1 | 244 | 1115 |
| 12 | 50 | 1.3 | 15.4 | 350 | 385 | 24.6 | 989 | 1409 |
| 12 | 62 | 1.8 | 21.3 | 308 | 339 | 21.8 | 876 | 1504 |
| 12 | 77 | 3.4 | 40.5 | 204 | 225 | 14.5 | 583 | 1946 |
| 12 | 100 | 5.3 | 48.2 | 172 | 189 | 12.2 | 491 | 1648 |
| 5 | 40 | 2.4 | 11.8 | 147 | 162 | 16.9* | 636* | 966 |
| 5 | 59 | 5.6 | 17.5 | 96 | 106 | 11.1* | 415* | 799 |
| 5 | 79 | 10.3 | 35.8 | 70 | 76 | 8.0* | 300* | 1075 |
| 5 | 89 | 12.3 | 31.6 | 65 | 72 | 7.5* | 283* | 822 |

[a] The data are averages obtained by Barney et al. and Arora et al.[4,5] [b] Average number of Kuhn segments in an elastically active chain is estimated based on Equation S1. [c] Force-free contour length $L = bN$. [d] Except for the last four rows with *, the fracture force $f_f$ for a polymer chain is considered to be 2.5 nN as the C-S bond is the weakest bond along the backbone, which is a preferential scission site.[11] For the last four rows with *, the fracture force $f_f$ for a polymer chain is considered to be 3.8 nN as the alpha C-C bond of the triazole group is the weakest bond along the backbone, which is a preferential scission site.[12]

**Dependence of force-displacement on the retraction of tree-like structures.** In the main text, we approximate the network continua as tree-like structures. Based on this approximation, we can plot the force-displacement dependence for the quasi-static unloading of the tree-like structures (**Figure 3b** red). In generations directly connected to the bridging chain ($g = \pm 1$), at least one strand will be part of the topological loop (**Figure 3a**). Instead of using Equation 3, we use the equation below to describe the force-displacement of the tree-like structures:

$$R_{tree}(f) = \sum_{|g|=1}^{n} R\left(\frac{f}{(z-2)(z-1)^{|g|-1}}\right)$$

$$= L \cdot \sum_{|g|=1}^{n} \mathcal{L}\left(\frac{\frac{f}{(z-2)(z-1)^{|g|-1}}}{f_s}\right)\left(1 + \frac{\frac{f}{(z-2)(z-1)^{|g|-1}}}{f_e}\right) \quad (S5)$$

The unloading curves of the tree-like structures with different parameters are plotted in **Figure S1** (unload from 5 nN). The default parameters are $f_s = 5.86$ pN, $f_e = 105$ nN, and $z = 4$, which give the black curves in **Figure S1** and the red curve in **Figure 3c**. The single-chain parameters $f_s$ and $f_e$ clearly do not significantly influence the unloading curve accoarding to Figure S1a&b. The functionality of the tree-like structures has larger impact on the retraction, but the curves are still not qualitatively different. This suggest the quasi-static unloading of the network continua are more of network properties that are not significantly influenced by the molecular mechanics of single chain.

This section aims to show that the network continuum consists of different network chains might be similar in their retraction mechanics under the quasi-static condition. Therefore, the energy released from the network continuum is largely dictated by the stretchability of the opened loop, since it controls how much the network continuum can retract before the next loop is opened. This gives the semi-quatitative relationship in Equation 5. Note that we are not attempting to conclude any practical solutions for how the network continuum retracts after a bridging chain breaks. We acknowledge this process is complex and related to the polymer dynamics.

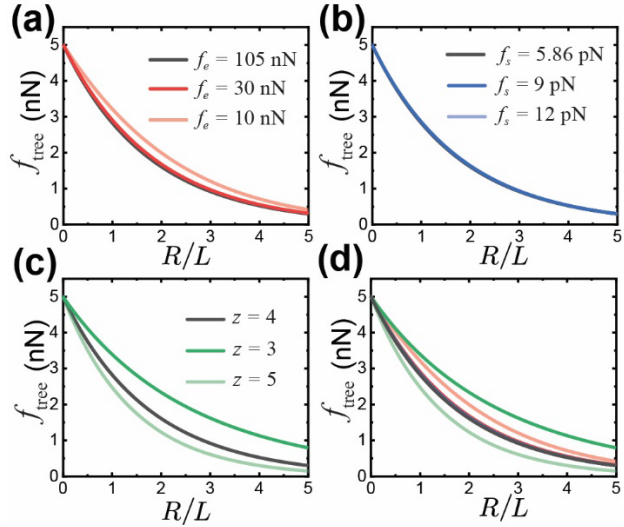

**Figure S1.** Force-displacement dependence for the retraction of tree-like structures from 5 nN with different single-chain parameters. The black curves in (a-d) are the same, and they have default parameters ($f_s$ = 5.86 pN, $f_e$ = 105 nN, and $z$ = 4) based on the experimental data for PEG polymers. (a) Change $f_e$ while keeping other parameters constant. (b) Change $f_s$ while keeping other parameters constant. (c) Change $z$ while keeping other parameters constant. (d) All curves in (a-c) plotted together.